# Utvärdering av effektivitet och kvalitet med strukturerade brottsanmälningar

*(Pre-print manuscript)*

Martin Boldt

*Institutionen för Datalogi och Datorsystemteknik*

*Blekinge Tekniska Högskola*

*371 79 Karlskrona*

*martin.boldt@bth.se*

**Abstrakt**

Denna artikel utvärderar en strukturerad metod för registrering av brottsplatsuppgifter från mängd-/vardagsbrott genom en jämförelse med traditionella textbaserade brottsanmälningar. En initial användarstudie kopplat till ett bostadsinbrott utvärderar effektiviteten hos de båda metoderna. Effektiviteten kvantifieras som dels tiden det tar att registrera bostadsinbrottet samt antalet relevanta brottsplatsuppgifter som registreras. Resultaten visar att den strukturerade metoden har statistiskt signifikant snabbare avrapportering än traditionella textbaserade anmälningar (p<0,05). Samtidigt registrerar den strukturerade metoden i snitt 2,96 gånger fler brottsplatsuppgifter. Utöver effektivitetsskillnaderna lyfts och diskuteras även mer kvalitativa för- och nackdelar med den strukturerade metoden. Sammantaget konstateras att metoden kan ge både stora tidsbesparingar och kvalitetsökningar i avrapporteringen vad gäller vardagsbrottens olika brottskategorier. En grov uppskattning landar kringsvid 11-30 heltidstjänster/år. Detta skulle kunna avlasta redan tungt belastade poliser, samt i viss mån frigöra resurser inom polismyndigheten som kan användas till mer kvalificerade uppgifter, t.ex. inom brottsamordningen.

*Nyckelord*: Brottsanmälningar, strukturerad brottsplatsuppgifter, rationell anmälningsrutin, effektivitetsutvärdering, SAB, RAR.

# 1  Introduktion

Av de brott som anmäls till polisen utgörs omkring 88% av så kallade vardagsbrott, även kallade vardagsbrott (Riksrevisionen, 2010). Vilket grovt räknat ger i häraden 1,3 miljoner brottsanmälningar under 2016. Vardagsbrott är ett samlingsbegrepp för flera olika brottskategorier vilka exempelvis inkluderar diverse tillgreppsbrott, skadegörelse, bedrägerier, misshandel och rattfylleri. De så kallade vardagsbrotten är av enklare beskaffenhet och drabbar årligen en betydande andel av befolkningen. Vardagsbrottens andel av den totala brottsligheten ser liknande ut även i andra länder. Enligt en uppskattning från 2004 var 78% (cirka 5,9 miljoner) brott i delar av England och Wales just vardagsbrott. Det är viktigt att inte bagatellisera brott bara för att de benämns som vardagsbrott. Dessa brott påverkar ofta målsägandens livssituation mycket negativ, särskilt för barn och äldre.

Eftersom vardagsbrotten är så pass frekvent förekommande och utgör en stor andel av samtliga brottsanmälningar är det intressant att undersöka förbättringsmöjligheter vad gäller både effektiviteten och kvaliteten i anmälningsrutinen för dessa brott, vilket också är syftet med denna studie. I Sverige upprättar polisen brottsanmälningar med datorprogrammet Rationell Anmälningsrutin (RAR) vilket infördes i början av 90-talet. RAR delar upp en brottsanmälan i olika sektioner där varje sektion sedan skrivs som löpande text i fritextformat. Trots att samtliga poliser utbildas i hur polisanmälningar upprättas i RAR, så medför brottsanmälningar skrivna i fritextformat viss godtycklighet vad gäller vilka uppgifter som samlas in från brottsplatser. En del uppgifter är obligatoriska, vid exempelvis bostadsinbrott registreras alltid adress, datum/tid och eventuella skador, medan det för andra uppgifter är mer öppet kring hur de dokumenteras, t.ex. relaterat till gärningsmannens modus operandi (MO) eller uppgifter rörande målsägaren och dennes bostad.

Att uppgifterna som ingår i RAR-anmälningar skiljer sig åt beror på att olika poliser gör olika iakttagelser och ställer olika frågor till målsäganden baserat på tidigare erfarenheter. Utöver att olika RAR-anmälningar innehåller olika typer av uppgifter så skiljer sig även de uppgifter som en specifik polis samlar in över tiden, t.ex. p.g.a. trötthet och stressnivå. Sammantaget kan man alltså konstatera att de brottsplatsuppgifter som samlas in vid ett brott inte alltid samlas in för ett annat brott, vilket resulterar i att brottsbeskrivningarna innehåller olika komponenter. Vilket innebär problem i ett senare skede då dessa brott behöver jämföras i brottsamordningsfunktionen. Som ett exempel på problemet kan vi tänka oss två inbrott (A och B) utförda av samma gärningsman som ännu är okänd för polisen. Gärningsmannen opererar genom att bl.a. ringa upp bostaden före inbrottet för att kontrollera om någon är hemma. Polispatrullen på plats vid inbrott A ställer frågan till målsäganden och dokumenterar att denne har ett okänt telefonnummer i listan över missade samtal strax innan inbrottet. Vid inbrott B ställer däremot en annan polispatrull aldrig den frågan till målsäganden. Därmed undanröjs också möjligheten för brottsamordningsfunktionen att senare kunna använda den här MO-detaljen (att gärningsmannen ringer upp bostaden för inbrottet) för att koppla samman inbrotten A och B i en potentiell serie. Genom att kombinera bevisen samtliga brottsplatser inom en serie så erhålls en tydligare bild av den eftersökta gärningsmannen, jämfört med att studera varje brott separat. Att så långt som möjligt utreda vardagsbrott i relevanta serier är en förutsättning för att dels använda polisiära resurser så resurseffektivt som möjligt (Woodhams et al., 2010), men även för att markant få upp personuppklarningsprocenten inom aktuella brottskategorier.

Exemplet ovan visar endast på en MO-detalj, men det finns så klart flera olika, t.ex. huruvida målsäganden parkerat sin bil på en flygplats, köpcentrum eller dylikt i samband med inbrottet. Ifall en uppsättning brott har *flera* sådana här MO-detaljer som stämmer överens, samtidigt som brotten visar intressanta relationer geografiskt såväl som i tid, så leder dessa indicier fram till hypotesen att dessa brott har utförts av samma gärningsman. Alltså, något konkret för polisen att utreda vidare. Men, eftersom RAR-anmälningarna generellt inte innehåller dessa uppgifter saknar också brottsamordningen underlag för att identifiera sådana likheter mellan brottsplatser.

För att hitta lösningar på dessa problem har polisen (främst UC Syd och utvecklingsavdelningen) sedan 2012 samverkat med Nationellt Forensiskt Centrum (NFC) och Blekinge Tekniska Högskola (BTH). Resultatet har mynnat ut i en strukturerad metod för insamling av brottsplatsuppgifter för följande tre vardagsbrottskategorier:
- inbrott i permanentbostad,
- drivmedelstölder, och
- transportstölder.

De två senare formulären har tagit fram av den Nationella Transportsäkerhetsgruppen. Den strukturerade metoden baseras på digitala formulär där poliser dokumenterar brottsplatser med hjälp av kryssrutor. Den främsta fördelen med denna metod är att poliserna har en inbyggd checklista över vilka uppgifter de *ska* samla in från brottsplatsen. Därmed samlas samma typ av uppgifter in från samtliga brottsplatser vilket gör att det finns en *minsta gemensam nämnare* med brottsplatsuppgifter. Denna minsta gemensamma nämnare av brottsplatsuppgifter möjliggör senare analyser för att hitta likheter mellan brott. Eftersom uppgifterna samlas in digitalt och kodas binärt finns det mycket goda möjligheter till automatiska analyser med hjälp av statistiska metoder såväl som självlärande algoritmer från området artificiell intelligens, t.ex. för automatisk gruppering av brott baserat på dess inbördes likhet.

Syftet med denna artikel är att utvärdera skillnaderna i effektivitet och kvalitet mellan traditionella RAR-anmälningar i fritext och den strukturerade metoden. Denna artikel presenterar därför en första initial användarstudie där den strukturerade metoden och traditionella RAR-anmälningar jämförs för brottskategorin inbrott i permanentbostad. Jämförelser görs med bäring på dels den kvantitativa effektiviteten såväl som för mer kvalitativa kvalitetsaspekter för respektive metod.

Dispositionen av artikeln är som följer, i nästkommande sektion ges en bakgrund där både den traditionella fritextbaserade metoden för brottsanmälningar och den strukturerade metoden beskrivs. Därefter beskrivs metodiken som använts i Sektion 3. Resultaten presenteras i Sektion 4 och därefter analyseras de och diskuteras i Sektion 5. Slutligen, presenteras slutsatserna i Sektion 6 följt av förslag till fortsatt forskning i Sektion 7.

## 2  Bakgrund

Denna sektion ger en bakgrundsbeskrivning av dels traditionella brottsanmälningar avseende bostadsinbrott såväl som för den strukturerade metoden. Dessutom belyses några nyckelutmaningar i polismyndighetens brottsamordningsfunktion.

### 2.1  Traditionell metod för registrering av brott

Processen för hur uppgiftsinsamling från bostadsinbrott går till skiljer sig åt baserat på hur pass nära inpå brottstillfället som polisen mottager en anmälan från målsäganden. Om en målsägande ringer om ett för stunden pågående bostadsinbrott högprioriteras ärendet och närmsta lediga patrull rycker ut. Om målsäganden å andra sidan upptäcker bostadsinbrottet vid hemkomsten från en semester kan polisen nedprioritera ärendet och besöka brottsplatsen vid senare tidpunkt, t.ex. nästkommande dag. På brottsplatsen gör patrullen iakttagelser samt talar med målsäganden. Detta görs dels för att samla in uppgifter men även för att förklara vad man bör tänka på för att undvika att röja eventuella spår på brottsplatsen. De iakttagelser som görs på platsen tillsammans med informationen från målsäganden dokumenteras i form av stödanteckningar i anteckningsblock, eller digitalt i mobila enheter eller i mobila polisdatorer. Om det finns möjlighet kallas även tekniker eller lokala brottsplatsundersökare till platsen för att genomföra en brottsplatsundersökning med syfte att säkra spår, exempelvis skoavtryck, verktygsspår och fingeravtryck. Slutdestinationen för all insamlad information är polisens Rationella Anmälningsrutin (RAR) i vilken en fritextrapport upprättas av patrullen på plats. En RAR-brottsanmälan består av oformaterad löpande text med den frivilliga möjligheten för avrapporterande polis att dela upp anmälan med rubriker. Detta är dock inget hårt krav i systemet utan det är upp till varje användare i respektive fall. För t.ex. ett bostadsinbrott delas RAR-anmälan oftast upp under följande fem rubriker: *Brottet*, *Omständigheter*, *Skador*, *Brottsplatsundersökning/spår*, och *Övrigt*. En RAR-anmälan kompletteras därefter av de som utfört brottsplatsundersökningen genom tilläggsanmälningar.

Det finns viss godtycklighet i vilka uppgifter som samlas in från brottsplatser och dokumenteras i RAR-anmälningar. En del uppgifter är obligatoriska, t.ex. adress och datum/tid, medan andra är mer öppna kring hur de ska dokumenteras, t.ex. vad gäller gärningspersonens modus operandi (MO), eller uppgifter rörande. Skillnaderna kring vilka brottsplatsuppgifter som samlas in beror på att olika poliser gör olika iakttagelser och ställer olika frågor till målsäganden, vilket resulterar i att olika brottsplatsuppgifter samlas in från olika bostadsinbrott. Därtill skiljer sig även de uppgifter som en specifik polis samlar in över tiden, t.ex. p.g.a. trötthet, stressnivå etc. Man kan alltså konstatera att de frågor som ställs vid ett inbrott inte alltid ställs vid ett annat vilket betyder att beskrivningarna av brotten innehåller olika komponenter. Vidare innebär brottsanmälningar skrivna som fritext problem, i olika utsträckning, p.g.a. felstavningar och att poliser dokumenterar samma brottsplatsuppgift på olika sätt då personer formulerar sig olika, t.ex. genom olika synonymer. Detta innebär problem då olika brott vid senare tillfälle ska jämföras i brottsamordningen.

Sammantaget gör detta att det blir omöjligt att på ett kvalificerat och effektivt sätt söka efter likheter i diverse brottsdetaljer mellan brott baserat på dess RAR-anmälningar. Framförallt eftersom 1) RAR-anmälningarna inte består av samma komponenter då de innehåller olika brottsplatsuppgifter, och i mindre utsträckning att 2) dessa har dokumenterats på olika sätt. Dessa utmaningar i brottsamordningen påverkar också personuppklarningsprocenten negativt. För bostadsinbrotten under 2016 var denna endast 3.5% (Brottsförebyggande rådet, 2017).

## 2.2 Strukturerad insamling av brottsplatsuppgifter

I ett antal forskningsprojekt har polisen, Nationellt Forensiskt Centrum (NFC) och Blekinge Tekniska Högskola (BTH) gemensamt samverkat för att adressera utmaningarna relaterade till RAR-anmälningar och samordningsproblemen som brottsamordningen stöter på genom ett fullt ut medarbetardrivet arbetssätt. Projekten har resulterat i en metod för att registrera brottsplatsuppgifter på ett mer strukturerat sätt än RAR-anmälningarna (Boldt, Borg, & Melander, 2015). Den strukturerade metoden används i skrivande stund i olika utsträckning för registrering av bostadsinbrott i södra Sverige, främst region Syd och Väst, där det har registrerats drygt 20 000 bostadsinbrott med metoden.

En förutsättning för att brottsamordningen bättre ska kunna länka samman brott i brottsserier är att *enhetliga* uppgifter från *samtliga* brottsplatser samlas in. Att samma frågor ställs till all målsägande och samma detaljer från brottsplatserna samlas in för samtliga brott inom en viss brottskategori är centralt för att senare effektivt kunna jämföra brotten sinsemellan. Vidare behöver de insamlade brottsplatsuppgifterna vara maskinläsbar (utan tolkningsutrymme) för att datorer på ett exakt och automatiskt sätt ska kunna utföra parvisa jämförelser av samtliga brott. En sådan datorstödd analys gör det möjligt för brottssamordningen att dra nytta av automatiska jämförelser av likheter mellan flera olika MO-detaljer för att inte bara identifiera lokala brottsserier, utan även regionala och nationella brottsserier.

*Figur 1: Återgivning av sektionen för "Typ av bostadsområde" från SAB-formuläret för en enskilt belägen bostad på landsbygden.*

Dessa förutsättningar har resulterat i en lösning som bygger på digitala kryssbaserade PDF-formulär där polisen kryssar för de kryssrutor som beskriver brottsplatsen. För att testa och utvärdera denna lösning valdes brottskategorin bostadsinbrott för vilken ett sådant digitalt kryssbaserat formulär, med namn *Standardiserad Anmälningsrutin Bostadsinbrott* (SAB), skapades. Innehållet i formuläret är baserat på polisens domänkunskap om brottskategorin bostadsinbrott. Formuläret, består av 11 sektioner vilka totalt innehåller 133 parametrar i form av kryssrutor samt ett antal textfält för diarienummer, adress, tidsangivelser och övriga anteckningar, se Tabell 1. Sist i SAB-formuläret finns även ett fritextfält där övriga iakttagelser kan dokumenteras. Som ett exempel finns en av de 11 sektionerna, den som preciserar Typ av bostadsområde, återgiven i Figur 1. Ifall ett inbrott skett i ett ensligt hus ute på landsbygden så registreras detta genom två kryss, ett i rutan för *Tätort* och ett i rutan för *Enskilt*. Vid osäkerhet kan man klicka på frågetecknet uppe till höger i figuren så visas en hjälptext som beskriver de olika kryssrutorna i sektionen. SAB-formuläret utgörs av en vanlig PDF-fil och ryms på två A4-sidor. Alla kryss-markeringar görs alltså digitalt i programmet Acrobat Reader.

| Sektionsnamn | Beskrivning | Parametrar |
|---|---|---|
| Brottsplats | Datum och tidsangivelser, ifall det är fullbordat eller försök | 12 |
| Typ av bostadsområde | Tätort eller landsbygd, samt tomtens beskaffenhet | 7 |
| Typ av bostad | Villa, gård, par/radhus eller lägenhet. Standard samt antal plan etc. | 12 |
| Larm | Ifall det finns larm och om det var aktiverat, utlöst eller saboterat | 5 |
| Brottspreventiva åtgärder | Brottspreventiva åtgärder som vidtagits, t.ex. tömt posten | 10 |
| Målsägande | Hemma eller borta under brottet, märkliga iakttagelser som gjorts | 16 |
| Ingång objekt | Gärningsmannens tillvägagångssätt för att ta sig in | 26 |
| Genomsök | Vilken typ av genomsök som gjorts i bostaden | 3 |
| Gods | Godskategorier som tillgripits, t.ex. läkemedel, guld/smycken etc. | 19 |
| Spår | Typer av spår som säkrats på platsen, t.ex. skoavtryck, DNA etc. | 18 |
| Övrigt | Om det finns vittnesuppgifter, gods märkt med MärkDNA etc. | 5 |
| **Totalt** | | **133** |

*Tabell 1: Summering av innehåll i formuläret för Standardiserad Anmälningsrutin Bostadsinbrott.*

Efter att SAB-formuläret fyllts i så registreras det i en central databas genom en knapptryckning. Innan formuläret registreras i den centrala databasen utförs en automatisk verifikationskontroll av uppgifterna i formuläret. Om någon uppgift inte fyllts i, eller fyllts i felaktigt, uppmärksammas polisen på detta och registringen avbryts. Först då den inbyggda verifikationsprocessen genomgåtts utan fel kan formuläret registreras. På så vis minskar risken för att den centrala databasen, som lagrar alla SAB-formulär, innehåller felaktiga uppgifter. När bostadsinbrottet väl lagrats i den centrala databasen sparas det i ett digitalt maskinläsbart format där parametrarna kodas binärt (1 för ett kryss i en ruta, annars 0).

Att samma typ av uppgifter samlas in från samtliga bostadsinbrott samt att de kodas binärt innebär att kvaliteten på de insamlade uppgifterna ökas signifikant jämfört med textbaserade anmälningar.

Det är alltså möjligt att i en dator beskriva ett brott som en vektor med 133st ettor och nollor, en för varje parameter i SAB-formuläret. Detta har stor betydelse om man vill utnyttja de möjligheter som datoralgoritmer kan erbjuda.

## 2.3   Utmaningar och möjligheter för brottsamordningen

Det faktum att olika uppgifter samlas in från brottsplatser, samt dokumenteras i fritext på olika sätt, innebär problem då man vill upptäcka brott som utförts på samma sätt. Eftersom inte samma uppgifter samlas in från samtliga brottsplatser så innebär det att brottsamordningen blir lidande då de i många fall saknar underlag för att knyta ihop serier baserat på likheter/mönster i uppgifterna mellan brottsplatserna. Det blir på sätt och vis som att jämföra äpplen med päron. Även om det finns datorprogram som automatiskt kan analysera fritextrapporter i jakt på likheter så kan dessa inte prestera bättre än den indata i form av t.ex. RAR-anmälningar de har att arbeta med. Vilka alltså saknar uppgifter som kan vara av betydelse. Därutöver kan synonymer och olika sätt att formulera sig utgöra problem för automatiserade datoralgoritmer baserade på fritextdata då dessa försöker hitta likheter mellan brott.

Eftersom det inte är möjligt för brottsamordnarna att genomföra kompletta parvisa jämförelser mellan t.ex. bostadsinbrott så krävs istället andra arbetssätt. Idag förekommer det därför att man inom brottsamordningen manuellt analyserar flödet av nya brott som kommer in, och i detta flöde försöker upptäcka mönster, t.ex. en gruppering av brott vars MO sticker ut vilket kan indikera att dessa ingår i en gemensam serie. Men eftersom den information som analyseras är i form av fritext kan man inte göra heltäckande historiska återblickar p.g.a. den tid det tar att manuellt läsa igenom RAR-anmälningarna. En kompromiss är istället att selektivt analysera de mönster som identifieras i flödet gentemot historisk data för att koppla samman brott. Ett stort problem med detta arbetssätt är att det missar många serier eftersom dessa inte sticker ut under den korta nulägesbild som analyseras löpande. Dessutom genomförs dessa analyser endast inom ett begränsad geografisk område så arbetssättet missar brottsserier som har stor geografisk spridning.

Registrering av brott med SAB-formuläret inkluderar en process som säkrar att samma typ av information samlas in på ett strukturerat sätt från olika brottsplatser. Genom att den insamlade informationen alltid kodas på samma sätt, t.ex. att gärningsmannens MO alltid registreras likadant, är det senare möjligt att utföra automatiska jämförelser av brott. På så vis är det möjligt för datorprogram att automatiskt söka igenom de insamlade brottsplatsuppgifterna med syfte att identifiera likheter mellan olika brott som kan indikera att dessa ingår i en gemensam serie. Datoralgoritmer kan alltså påvisa möjliga samband mellan brott som sedan brottsamordnare inom polisen behöver analysera och värdera baserad på sin erfarenhet för att slutligen antingen förkasta eller acceptera påstådda serier. Detta leder till att polisen på ett effektivare sätt kan knyta liknande brott till nya serier utförda av gemensamma gärningsmän. Genom att i större utsträckning kunna fälla gärningsmän för serier av bostadsinbrott snarare än enstaka brott är förhoppningen att öka uppklarningsprocenten inom brottskategorin. På så vis adresserar man också den organiserade brottsligheten som ligger bakom en försvarlig andel av bostadsinbrotten.

Det är viktigt att understryka att dessa automatiserade metoder för brottsamordning på *inga* sätt är tänkta att ersätta polisens yrkeskunnande. De kommer istället utgöra hjälpmedel som kan indikera var sannolika länkar mellan enskilda brott finns, vilket i sin tur den mänskliga brottsamordnaren kan analysera och värdera. Genom dessa indikationer på länkar mellan brott ökar möjligheten för att koppla samman viktiga informationsfragment som är spridda mellan flera brottsplatser vilka kan skilja stort vad gäller både geografisk och tidsmässig utbredning. Ju fler informationsfragment från olika brottsplatser som kopplas samman i relevanta serier desto bättre eftersom detta dels ökar förståelsen för hur gärningsmännen beter sig inom brottskategorin, men också för att det ökar sannolikheten att finna tillräckligt med indicier och bevis för att nå personuppklarning. Dessutom möjliggör en gemensam undersökning av flera brott en mer effektiv användning av brottsbekämpande resurser (Woodhams et al., 2010).

De automatiska metoderna för brottsamordning ska användas som rena *interna* polisiära hjälpmedel i form av selekteringsverktyg som låter polisen selektera i sina stora informationsmängder och därmed hitta nålarna i hö-berget av inkommande ärenden. De är alltså inte tänkta att användas som underlag i t.ex. Tingsrättsförhandlingar etc. Avslutningsvis, i detta arbete fokuseras på brottskategorin bostadsinbrott, men den strukturerade metodens fördelar kan likaväl användas på andra brottskategorier av hög frekvens som involverar seriebrottslighet, t.ex. bedrägerier, åldringsbrott och IT-brott.

# 3 Metod

Denna sektion beskriver designen av studien, den föregående statistiska styrkeanalysen som genomfördes samt vilka utvärderingsmetoder som använts.

### 3.1.1 Experimentdesign och variabeluppsättning

För att studera eventuella skillnader i effektivitet mellan RAR-anmälningar och den strukturerade metoden vad gäller insamling av brottsplatsuppgifter så genomfördes en experimentell användarstudie. Studien mätte kvantitativt hur lång tid (i sekunder) det tog att beskriva ett fiktivt inbrott i dels en RAR-anmälan samt med SAB-formuläret. Dessutom mättes hur många unika brottsplatsuppgifter som respektive metod registrerade. Utöver de båda kvantitativa måtten analyserades också ett antal kvalitativa kvalitetsaspekter för respektive metod. Som grund för studien användes ett fiktivt brottsfall avseende ett inbrott i permanentbostad, vilket tillhandahölls av NFC. Det fiktiva brottsfallet används av NFC i den nationella utbildningen av kriminaltekniker och lokala brottsplatsundersökare.

Datainsamlingen under användarstudien skedde med hjälp av en webbaserad enkät och en experimentell metod valdes för att undersöka skillnaderna mellan registreringsmetoderna. Den oberoende variabeln är registreringsmetoden, och variabeln har följande två nivåer: registrering i fritext (RAR) respektive med det strukturerade SAB-formuläret (SAB). De beroende variablerna utgörs av 1) tiden det tar att registrera brottsfallet med respektive metod (mätt i sekunder), och 2) antalet unika brottsplatsuppgifter som respektive metod registrerar. En inom-gruppslig experimentdesign användes (engelska *within-subjects* eller *repeated measures*) och ordningen i vilken deltagarna registrerade det fiktiva brottsfallet med respektive metod skiftades för att motverka eventuella inlärnings- och utmattningseffekter (Shadish, Cook, & Campbell, 2002).

### 3.1.2  A priori styrkeanalys och deltagare

För att uppskatta antalet försökspersoner till studien så utfördes en a priori statistisk styrkeberäkning i statistikverktyget R med hjälp av tilläggspaketet *pwr* (Sheshkin, 2011). Den antagna effektstorleken mellan RAR och SAB uppskattades enligt Cohen's $d$ till 0,8 baserat på diskussioner med medarbetare hos polisen. En statistisk signifikansnivå på 0,05 ansågs vara lämpligt då arbetet utgör en initial studie. Vidare specificerades ett tvåsidigt parat test med en lägsta accepterad statistisk styrka på 0,8. Detta resulterade i ett uppskattat antal deltagare av 14 stycken.

Poliserna som deltog i studien var samtliga verksamma som poliser inom polisens region syd. De valdes ut genom att en medarbetare på polisens UC Syd i Malmö skickade en öppen förfrågan till två lokalpolisområden i södra Skåne. Alltså användes *convenience sampling*, d.v.s. tillgängliga poliser medverkade i studien (Robson, 2002). Totalt deltog 19 poliser i studien, men av dessa var fem stycken tvungna att utelämnas. Två stycken på grund av att de inte skickade in någon fritext-beskrivning av brottet, och ytterligare tre stycken för att de lämnade in helt tomma SAB-formulär. Resultaten i studien är baserade på de kvarvarande 14 deltagarna. Deltagandet i studien var frivilligt och anonymt.

### 3.1.3  Statistiska metoder

För att utvärdera eventuella statistiska skillnader mellan RAR och SAB med avseende på inrapporteringstid respektive mängd insamlad brottsplatsuppgifter så användes två stycken parade tvåsidiga *t-test* vid signifikansnivån 0,05 (Walpole, Myers, & Myers, 2014). Det parametriska t-testet valdes eftersom mätresultaten i de båda mätningarna kunde antas uppfylla antagandena för testet, t.ex. att de är någorlunda normalfördelade. För att kvantifiera den praktiska signifikansen i skillnaden mellan RAR och SAB så användes Cohen's $d$ (Cohen, 1988). Medan det vanliga p-värdet anger hur sannolikt det är att erhålla en viss uppsättning mätvärden, så anger istället Cohen's $d$ hur stor skillnaden är mellan grupperna. Det är därför intressant att komplettera p-värdet med Cohen's $d$ som ett mått på den praktiska signifikansen, även kallad effektstorleken, då detta mått kvantifierar skillnaden mellan de båda grupperna.

Alla mätvärden i denna artikel summeras med medelvärden tillsammans med standardavvikelser som ett mått på dess interna spridning. Dessutom visualiseras samtliga mätvärden med box-plottar som tydligt återger varje enskilt mätvärde såväl som kvartiluppdelningen bland mätvärdena. All statistisk analys utfördes i R med hjälp av erforderliga tilläggspaket.

# 4 Resultat

Resultaten presenteras i tre steg, först mätvärdena för hur lång tid respektive metod behövde för att registrera brottsplatsen. Därefter presenteras mätvärdena för hur många brottsplatsuppgifter som respektive metod registrerade. Slutligen beräknas den sammanvägda effektivitetsskillnaden baserat på de båda tidigare aspekterna.

## 4.1 Inrapporteringstid

I medelvärde behövde deltagarna 793 sekunder på sig för att registrera brottsplatsen med traditionell fritext (RAR). Alltså, drygt 13 minuter. Standardavvikelsen var 512 sekunder vilket visar att det fanns relativt stor variation i tid mellan deltagarna. För SAB var inrapporteringstiden i snitt 478 sekunder med en standardavvikelse på 154 sekunder. Skillnaden i medelvärde var 315 sekunder, eller drygt 5 minuter. Det betyder att givet deltagarna i denna studie var RAR 66% långsammare än SAB. Dessutom var standardavvikelsen för SAB drygt tre gånger lägre än för RAR, vilket betyder att inrapporteringstiderna för SAB var mer samlade.

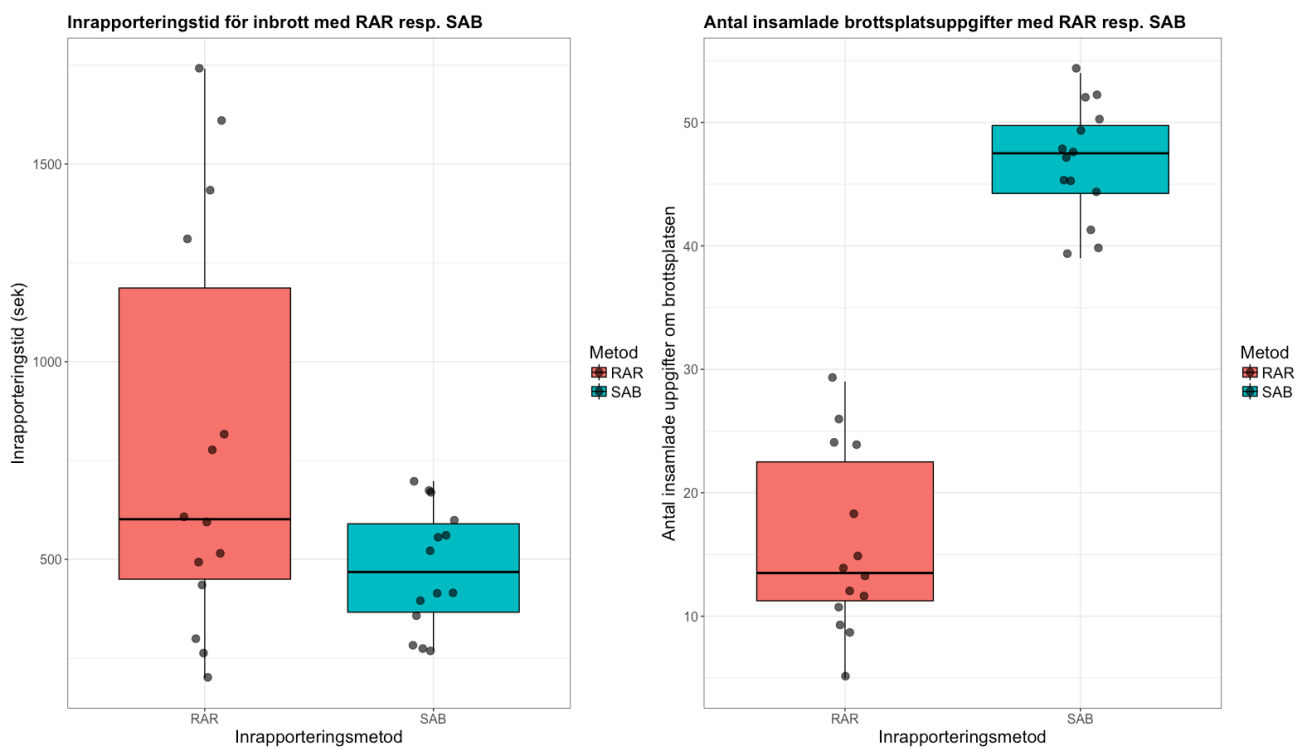

*Figur 2: Box-plotten till vänster visar inrapporteringstiden i sekunder för dels den fritextbaserade metoden (RAR) respektive den strukturerade metoden (SAB). Box-plotten till höger visar antalet registrerade brottsplatsuppgifter för de båda metoderna.*

Till vänster i Figur 2 visas inrapporteringstiden för respektive metod. De heldragna linjerna inuti boxarna visar medianerna (eller andra kvartilerna) medans under- och överkant visar första- respektive tredjekvartilerna. I figuren är det tydligt att mediantiden för RAR är något högre än för SAB samt att den senare har en betydligt mer samlad uppsättning mätvärden. RAR har specifikt fyra stycken mätvärden som drar upp inrapporteringstiden, vilket tas upp och diskuteras senare.

T-testet visade på signifikanta skillnader mellan de båda metoderna med SAB som den snabbare kandidaten ($p < 0{,}05$; $t=2{,}18$). Kvantifieringen av effektstorleken, eller den praktiska signifikansen, mellan de båda metoderna gav ett Cohen's *d* på 0,82. Vilket givet de allmänna riktlinjerna för tolkning av effektstorleken kan summeras som en någorlunda stor skillnad i tidsåtgång mellan metoderna (Cohen, 1988).

### 4.2   Mängd insamlad brottsplatsuppgifter

I medelvärde registrerade deltagarna 15,79 brottsplatsuppgifter med RAR. Standardavvikelsen var 7,00 brottsplatsuppgifter vilket visar att det fanns variation i antal insamlade uppgifter mellan deltagarna. För SAB var antalet registrerade brottsplatsuppgifter i snitt 46,71 stycken med en standardavvikelse på 5,02 uppgifter. SAB har alltså betydligt högre antal registrerade brottsplatsuppgifter i medelvärde samtidigt som dess interna variation är lägre jämfört med RAR. Skillnaden i medelvärde visar att SAB registrerar 2,96 gånger fler brottsplatsuppgifter än RAR.

Box-plotten till höger i Figur 2 visar antalet registrerade brottsplatsuppgifter för respektive metod. Baserat på box-plotten är det tydligt att antalet registrerade brottsplatsuppgifter är betydligt högre för SAB än för RAR, samt att den förra har en något mer samlad uppsättning mätvärden med lägre varians. De fyra mätvärdena för RAR som drog upp dess inrapporteringstid återfinns även i topp i box-plotten över antalet registrerade brottsplatsuppgifter. Dessa fyra deltagare i studien använde alltså mer tid än övriga deltagare då de registrerade brottsplatsen med RAR, men samtidigt registrerade de också fler brottsplatsuppgifter med RAR än övriga deltagare. I nästa sektion jämförs effektiviteten mellan de båda metoderna genom att bland annat jämföra hur lång tid det i snitt tog att registrera en brottsplatsuppgift med respektive metod.

T-testet visade på signifikanta skillnader mellan de båda metoderna, med SAB som den snabbare kandidaten ($p < 0{,}05$; $t=-13{,}40$). Vid analys av den praktiska signifikansen mellan de båda metoderna så erhölls ett Cohen's *d* på imponerande 5,01. Vilket kan summeras som en mycket stor skillnad i antalet registrerade brottsplatsuppgifter mellan SAB och RAR.

### 4.3   Sammanvägd effektivitetsskillnad

För att jämföra effektiviteten mellan SAB och RAR så beräknades hur lång tid det i snitt tog för respektive metod att registrera en brottsplatsuppgift. Detta beräknades genom att dividera medeltiden för respektive metod med medelantalet brottsplatsuppgifter som metoden registrerade.

För RAR innebar detta att det i snitt krävs 50,21 sekunder per registrerad brottsplatsuppgift, medan det tog 10,22 sekunder per registrerade brottsplatsuppgift för SAB. Sammantaget innebär detta att SAB är 4,91 gånger mer effektiv än RAR vid registrering av brottsplatsuppgifter i denna studie. Skillnaden är statistiskt signifikant ($p < 0,05$).

# 5 Diskussion

De kvantitativa resultaten ovan visar tydligt att den strukturerade metoden för registrering av inbrott är mer effektiv än fritext-registreringar när hänsyn tas till både inrapporteringstiden och antalet registrerade brottsplatsuppgifter. Dessutom är det rimligt att anta att ytterligare tidsvinster är möjliga med den strukturerade metoden ju oftare personalen använder den. Generellt gäller att ju fler gånger och oftare en person genomför en viss uppgift desto snabbare går det att slutföra den, upp till en viss nivå då det planar ut. För den strukturerade metoden innebär det att ju fler gånger och oftare metoden används desto mer effektiva blir användarna i sin användning av den. Detta eftersom poliserna efterhand får en mer detaljerad mental bild av processen. Det är viktigt att understryka att även om en person aldrig, eller mycket sällan, använt den strukturerade metoden så är den ändå tidseffektiv jämfört med fritext-registreringar.

## 5.1 Kvalitativa kvalitetsvinster

Utöver de klara effektivitetsvinsterna medför den strukturerade metoden även mer kvalitativa kvalitetsvinster. Nedan listas dessa för att därefter individuellt diskuteras i mer detalj.

1. *förbättrad möjlighet för styrning och ledning vid brottsregistreringen,*
2. *obligatorisk check-lista för personal på brottsplatser,*
3. *registrerade brottplatsuppgifter samlas in binärt vilket ger helt nya sök- och analysmöjligheter (jämfört med fritext-registrering),*
4. *enhetligt registrerade brottsplatsuppgifter för bättre brottsamordning,*
5. *möjlighet till automatisk generering av fritext utifrån de strukturerade uppgifterna,*
6. *minutoperativ registrering av brottsplatsuppgifter,*
7. *klassificering av gärningsmän baserat på aggregerade brottsplatsuppgifter,*
8. *en rigorös registreringsprocess som signalerar positiva värden till målsäganden.*

Förbättrad möjlighet för s*tyrning och ledning vid brottsregistreringen* möjliggörs genom att polisledningen i den strukturerade metoden kan precisera vilka brottsplatsuppgifter som är obligatoriska att registrera. Då en polis registrerar ett brott med den strukturerade metoden valideras datan i formuläret automatiskt och eventuella ofullständigheter måste hanteras innan registreringen kan slutföras. Detta möjliggör en styrnings- och ledningsfunktion som låter säkerställa att obligatoriska brottsplatsuppgifter alltid registreras med den strukturerade metoden. Det behövs alltså inte några extra lathundar, checklistor eller dylikt för att instruera poliserna.

Avslutningsvis finns det även en hjälptext för varje sektion i det strukturerade formuläret som ger hjälp ifall någon oklarhet uppstår.

Det strukturerade formuläret inkluderar per automatik en *inbyggd check-lista* för poliser på brottsplatsen. Genom att följa den strukturerade metoden säkerställs att fler relevanta brottsplatsuppgifter från brottsplatsen registreras jämfört med fritext-registrering som snarare utgår från vad polisen vid det givna tillfället kommer ihåg att samla in baserat på utbildning och tidigare erfarenheter. Dessutom kan stress och trötthet negativt påverka den traditionella fritext-registreringen eftersom den är mer kognitivt krävande. Ifall polisen på plats gör iakttagelser som inte täcks upp av den strukturerade metoden så finns det ett öppet anteckningsfält i slutet av formuläret i vilket det går att dokumentera kompletterande uppgifter.

Med den strukturerade metoden *samlas alla brottsplatsuppgifter in binärt* eftersom ett kryss i en ruta kodas som 1 och uteblivet kryss kodas som 0. Detta gör att datorprogram enkelt kan ges en detaljerad och mycket exakt representation av ett brott genom att mata den med en uppsättning ettor och nollar. Genom att ge datorprogram sådana detaljerade beskrivningar över brott kan de också automatiskt leta efter likheter mellan brotten med hjälp av s.k. klustringsalgoritmer. Vidare är det möjligt att träna upp andra datorprogram (s.k. klassificeringsalgoritmer) på de brott som man vet har utförts av en viss gärningsman genom länkning baserat på fingeravtryck eller DNA. Därefter kan klassificeringsalgoritmerna prediktera vilka av alla andra brott som är mest sannolika att höra till respektive serie, baserat på hur lika de är i sitt modus operandi (Reich & Porter, 2015) (Yokota & Watanabe, 2002). Inget av detta skulle vara möjligt att göra på lika hög detaljnivå med textbaserade RAR-anmälningar. Framförallt eftersom den registrerar signifikant färre brottsplatsuppgifter, men även eftersom uppgifterna samlas in som text vilket är svårare för ett datorprogram att tolka, t.ex. på grund av felstavningar, synonymer och ofullständig dokumentering.

Genom den strukturerade insamlingen av brottsplatsuppgifter säkerställs att en *enhetlig bas med jämförbara brottsplatsuppgifter* samlas in från samtliga brott inom en viss brottskategori. Detta medför att man i ett senare skede i brottsamordningsfunktionen kan jämföra uppgifter mellan olika brottsplatser baserat på denna minsta gemensamma nämnare av brottsplatsuppgifter. För ett bostadsinbrott samlas t.ex. alltid in om bostaden var enslig belägen eller hade flera grannar, exakt hur gärningsmannen tog sig in, var han tog sig in osv. Vid fritextregistrering missas många sådana detaljerade brottsplatsuppgifter eftersom polisen som skriver brottsanmälan omöjligt kan komma ihåg samtliga relevanta brottsplatsuppgifter per brottskategori som ska registreras. Initiativ med separata check-listor som ska hjälpa poliserna med detta riskerar att komma bort efter en tid, vilket gör att de inte är uthålliga. I slutändan är det den enskilda polisen som skriver brottsanmälan. Avsaknaden av enhetliga brottsplatsuppgifter från samtliga brottsplatser försvårar brottsamordningsfunktionen eftersom det inte lika precist går att undersöka likheter och mönster mellan olika brottsplatser baserat på textanmälningar, som det går för de enhetligt och detaljerade brottsplatsuppgifterna. Eftersom den strukturerade metoden samlar in brottsplatsuppgifterna binärt kan datorprogram bistå i identifieringen av intressanta mönster och likheter baserat på gärningsmannens modus operandi. Upptäckta likheter och mönster presenteras för brottsamordnare som därefter kan analysera dem mer utförligt. Det centrala är att brottsamordnarna inte längre manuellt behöver läsa igenom en stor mängd RAR-anmälningar i sina försök att hitta likheter mellan brott. De kan istället lägga sin tid det nästkommande steget i analysen, alltså att undersöka de redan identifierade mönstren i mer detalj. Detta innebär också att mönster kan upptäckas mellan vitt skilda geografiska platser såväl som över längre tidsperioder, t.ex. ett antal månader.

Binärt kodad data är enklast för en dator att behandla, men detta gäller inte för oss människor. Vi föredrar istället skriven text framför en mängd ettor och nollor då vi ska inhämta ny kunskap. Från de strukturerade brottsplatsbeskrivningarna för brott går det att automatiskt *generera fullständiga textbeskrivningar av brotten*. Detta genom ett datorprogram som gör om varje kryss till en separat mening, eller en del i en större mening. Därefter sammanställer programmet samtliga meningar till en enhetligt formaterad rapport som är rätt- och avstavad. Det går alltså utmärkt att använda den strukturerade metoden och samtidigt erhålla skrivna textrapporter. Som en bonus kommer dessa textrapporter vara 100% enhetligt formaterade, rätt-/avstavade och uppskattningsvis innehålla i snitt 2,96 gånger fler brottsplatsuppgifter än de traditionella fritextrapporterna. En prototyp av denna funktion finns implementerad i beslutsstödsystemet SAMS för att visa på möjligheten och styrkan i en sådan funktion.

Med den strukturerade metoden går det snabbt för första polis på en brottsplats att fylla i en brottsplatsbeskrivning och registrera denna i polisens gemensamma datalager. Detta kan ske redan inom några minuter vilket möjliggör *minutoperativa sökning och detektering av pågående brottstrender* genom automatisk spårning av MO-likheter, såväl som spatio-temporala samband bland registrerade brott. Tidigare forskning visar att tillgång till fler MO-detaljer ökar träffsäkerheten jämfört med att endast länka samman brott i serier baserat på spatio-temporala samband i geografi och tid (Bouhana, Johnson, & Porter, 2016) (Tonkin, Santtila, & Bull, 2011) (Bennell & Jones, 2005). Det centrala här är att dessa analyser kan göras betydligt snabbare med den strukturerade metoden eftersom en första registrering av brottsplatsen kan göras snabbt och korrekt. Därefter kan datoralgoritmer automatiskt och outtröttligt kontinuerligt leta efter mönster bland de brotts som rapporteras in i binärt format. Därmed ökar också möjligheten att identifiera *pågående* serier såväl lokalt som regionalt och nationellt.

De detaljerade brottsplatsuppgifterna för exempelvis bostadsinbrott innehåller ledtrådar kring gärningspersonens beteende, t.ex. hur riskbenägen personen är, eller i vilken utsträckning personen förberett sig för brottet. Ett sätt att dra nytta av dessa ledtrådar är genom att aggregera brottsplatsuppgifterna med hjälp av självlärande datoralgoritmer från ämnesområdet *machine learning*. Under 2016 genomfördes en sådan första initial studie tillsammans med profilerare från polisens Operativa Analysgrupp i Stockholm (Boldt, Borg, Svensson, & Hildeby, 2017).

Resultaten indikerar att automatisk aggregering av brottsplatsuppgifter till beskrivningar av beteenden hos gärningsmän är möjligt för bostadsinbrott. Samt att sådana aggregerade värden kan användas vid länkningen av brott både inom en brottskategorier, t.ex. inbrott, men även *mellan* olika brottskategorier, t.ex. för att länka inbrott mot drivmedelstölder baserat på gärningspersonens beteende.

Avslutningsvis signalerar den strukturerade metoden *positiva värden till målsäganden* då polisen tar upp en anmälan för ett brott. Genom att registrera en mängd relevanta brottsplatsuppgifter, både genom iakttagelser på brottsplatsen och genom frågor till målsäganden, så erhålls en detaljerad bild av brottet. Detta i sin tur signalerar positiva värden till målsäganden då denna känner att polisen engagerar sig i det aktuella brottsärendet och målsägandes situation.

### 5.2  Negativa invändningar mot den strukturerade metoden

Under utvecklingen av den strukturerade metoden har det även framkommit ett antal negativa invändningar. Dessa invändningar har kommit till vår kännedom dels då vi besökt lokalpolisområden runt om i Sverige för att informera om den strukturerade metoden, men även genom den representantgrupp med poliser från respektive län som i olika utsträckning använt SAB-formuläret. Nedan listas dessa invändningar för att därefter individuellt diskuteras i mer detalj.

1. *alltför uppstyrd metod,*
2. *"inte ännu ett formulär",*
3. *risk för att poliser inte fyller i formulären (korrekt),*
4. *formuläret innebär merarbete för den enskilda polisen.*

Vid några tillfällen har vi fått återkoppling kring att den strukturerade metoden är *alltför uppstyrd* jämfört med att skriva fritextrapporter. Sådan återkoppling har varit som en initial negativ invändning då SAB-formuläret hölls på att presenteras för poliser. I samtliga dessa fall har polisen som fällt kommentaren själv i efterhand påpekat att fördelarna med metoden vida överstiger nackdelarna. Det är viktigt att komma ihåg att den strukturerade metoden fortfarande möjliggör för polisen på brottsplatsen att ta egna initiativ och samla in ytterligare brottsplatsuppgifter med hjälp av anteckningsfältet sist i formuläret. Alltså genom kompletterande fritextdokumentering av brottet. Även om arbetet i denna studie beskriver ett PDF-formulär är det är fullt möjligt att istället utveckla en interaktiv app som poliserna kan använda på läsplattor eller smartphones. En sådan interaktiv app kan vara än mer lättanvänd jämfört med formulären.

Vid något tillfälle har vi även fått reaktionen, *"inte ännu ett formulär"* från poliser. Precis som vid klagomålen på den uppstyrda metoden, så har även dessa personer själva uttryckt hur enkel och lättanvänd den strukturerade metoden är efter att de fått prova den. Dessutom har det betonat hur mycket användbart det går att göra med de registrerade strukturerade brottsplatsuppgifterna, vilket inte går att göra baserat på RAR-anmälningarna. Vid lanseringen av en nationell modell för registrering av strukturerade brottsplatsuppgifter vore det därför bra att belysa nyttan av de insamlade uppgifterna för dels brottsamordning men även för möjligheten att utvinna ny kunskap om brottskategorin. Dessutom bör man så klart framföra tidsbesparingen för enskilda poliser som ett argument till införandet för att ytterligare öka motivationen hos poliserna.

Det har även kommit frågor rörande *risken att poliser inte fyller i formulären (korrekt)*. Även om detta är en risk så har vi inte sett några sådana problem för de drygt 20 000 formulär som samlats in för bostadsinbrott. En jämförelse för medelantalet kryss som fyllts i SAB-formulären under 2014 - 2016 visar det dessa legat konstant. Det finns alltså inte heller någon tendens till att poliserna fyller i färre brottsplatsuppgifter efterhand som åren går. Så, även om det alltid finns risk för att enskilda medarbetare inte fyller i formulären korrekt går det inte att se några strukturella problem.

Avslutningsvis har flera poliser framfört att SAB-formuläret innebär *merarbete* eftersom man ska fylla i *både* ett strukturerat formulär och en traditionell fritextanmälan för bostadsinbrott. Under utvecklingen av den strukturerade metoden har det inneburit ett visst merarbete för enskilda poliser då SAB-formuläret ska fyllas i samtidigt som obligatoriska RAR-anmälningar. Dock har detta inneburit att det samlats in betydligt fler relevanta brottsplatsuppgifter för de brott som även registrerats med SAB-formläret. Vid ett nationellt införande av den strukturerade metoden fyller poliserna endast i de strukturerade brottsplatsuppgifterna. Korrekta fritextrapporter för brotten automatgenereras därefter från den strukturerade data. Sammantaget sparar den enskilda polisen tid samtidigt som betydligt fler relevanta brottsplatsuppgifter samlas in.

## 5.3   Sammantagen effektivitetsvinst för hela polismyndigheten

Resultatet i denna studie indikerar att den strukturerade metoden kan spara drygt fem minuter per inbrottsanmälan jämfört med RAR-anmälningar. Om denna tidsbesparing skalas upp till de ca. 22 000 bostadsinbrott som anmäldes under 2016 erhålls 1 833 timmar, eller drygt en heltidstjänst á 1800 timmar/år. Detta gäller endast för brottskategorin bostadsinbrott. Att göra en liknande uppskattning för samtliga vardagsbrott är vanskligt då flera viktiga parametrar är okända, t.ex. hur många brottsplatsuppgifter det i normalfallet förkommer i respektive brottskategori och hur de relaterar till antalet för bostadsinbrott. Om man för enkelhetsskull skalar upp tidsbesparingen om fem minuter för bostadsinbrott för samtliga 1,3 miljoner vardagsbrott som anmäldes under 2016 erhålls en tidsbesparing på 108 333 timmar, eller drygt 60 heltidstjänster. Då detta med största sannolikhet är alltför högt räknat är det rimligare att resonera kring en halvering av tidsbesparingen för flera av brottskategorierna. Vilket innebär att man hamnar någonstans upp till 30 heltidstjänster. Medarbetare vid polisens Utvecklingscentrum Syd uppskattar att den strukturerade metoden ger en besparing på minst 11 heltidstjänster/år om den implementeras för följande brottskategorier:

åldringsbrott och brott mot funktionshindrade[1], transportstölder, Dieselstölder, bedrägerier, IT-brott[2], och bostadsinbrott. Så om den strukturerade metoden används för de brottskategorier som inkluderas i vardagsbrotten, t.ex. inbrott i butiker eller fritidshus, rån respektive skadegörelse, så är det rimligt att den totala besparingen inom polismyndigheten hamnar kringsvid 11-30 heltidstjänster/år.

Som nämnts tidigare är det dessutom troligt att räkna med än högre tidsbesparing p.g.a. inlärningseffekter hos den enskilde polisen, ju oftare den strukturerade metoden används. Vidare finns möjlighet att vidareutveckla metoden ytterligare genom att gå från dagens PDF-formulär till förmån för interaktiva appar istället. Även detta har potential att öka effektiviteten och göra metoden än mera lättanvänd för polisen så besparingen hamnar i den övre delen av skalan. Det skulle även vara intressant att undersöka effekterna av användandet av strukturerade anmälningar inom polisens kontaktcentrum (PKC) som hanterar stora flöden inkommande brottsanmälningar via telefon. Som utgångspunkt är det rimligt att anta att det finns stora effektiviseringsmöjligheter även här. Vidare skulle det även vara intressant att studera effekterna av strukturerade anmälningar på kö-längderna hos PKC.

Utöver rena tidseffektiviseringar inom polisen så tillkommer även fördelarna med att betydligt fler brottsplatsuppgifter samlas in jämfört med tidigare. Samt att de insamlade brottsplatsuppgifterna är av högre kvalitet p.g.a. såväl de inbyggda styrnings- och ledningsfunktionerna genom den automatiska valideringsrutinen som måste godkännas innan en brottsplatsbeskrivning kan registreras. För att fullt ut dra nytta av dessa fördelar behövs anpassade analyssystem för de insamlade brottsplatsuppgifterna, vilket också diskuteras härnäst.

### 5.4  Analysstöd för effektivare och kraftfullare brottsamordning

För att på bästa sätt utnyttja de detaljerade brottsplatsbeskrivningarna som den strukturerade metoden erbjuder behövs en ny paradigm polisiära analysmetoder inom polisen som kan dra nytta av de nya stora datamängderna med brottplatsuppgifter, cf. BigData. Sedan starten 2012 har polisen, NFC och Blekinge Tekniska Högskola arbetat medarbetardrivet och agilt för att ta fram prototyper till sådana analysmetoder. Denna samverkan har mynnat ut i ett analyssystem kallat SAMS (Strukturerat Analysverktyg för Mängd- och Seriebrott) som utvecklats vid Blekinge Tekniska Högskola, men som kostnadsfritt kan användas och vidareutvecklas av polisiära organisationer. Analyssystemet SAMS är webb-baserat vilket innebär att alla sökningar och analyser görs i valfri webbläsare. SAMS har främst följande tre styrkor:
1. *effektiva och lättanvända sökningar* bland vardagsbrott som möjliggör sökningar baserat på en stor mängd MO-parametrar,

---

[1] Exklusive vinningsbrott inom ramen av bedrägerier.
[2] Också exklusive vinningsbrott inom ramen av bedrägerier.

2. *avancerade och kompetenta analyser* för att hitta samband mellan de registrerade brottens brottsplatsuppgifter,
3. *möjlighet att automatiskt göra sökningar och analyser i bakgrunden* och rapportera tillbaka intressanta resultat.

Ett exempel på en *effektiv sökning* bland inrapporterade bostadsinbrott är exempelvis då en brottsamordnare använder sökfunktionen i SAMS och kryssar i följande sex kryss på söksidan:
- villa,
- ensligt belägen,
- målsäganden hemma under brottet,
- ingång genom olåst fönster,
- försiktigt genomsök, och
- stöld av kontanter och smycken.

Att både precisera och genomföra en sådan sökning tar med SAMS mindre än en minut, men att genomföra samma sökning m.h.a. traditionella textbaserade RAR-anmälningen skulle vara omöjligt då dessa många gånger saknar flera av de nödvändiga informationskomponenterna, t.ex. om bostaden var ensligt beläget eller om GM genomfört ett försiktigt genomsök. Även om textbaserade RAR-anmälningar skulle ha inkluderat denna information skulle det ändå vara näst intill omöjligt att utföra en korrekt sökning m.h.a. sökord som exempelvis "ensligt" eftersom en synonym till ordet kan ha använts, t.ex. avsides eller öde, alternativt att ordet är felstavat. Om brottssamordnaren därefter skulle vilja modifiera sökningen, t.ex. genom att även inkludera ingångsmoduset att borra upp lås, så görs detta på färre än 30 sekunder i SAMS. Att göra samma justering av sökningen baserat på textanmälningar kräver upp emot lika mycket tid som det tog att göra den första sökningen. Dessa lättanvända och snabba sökningar i SAMS möjliggör ett både iterativt och interaktivt arbetssätt som gör att polisen kan utvinna mycket ny kunskap om brottslighet.

Vidare inkluderar SAMS även en plattform för att använda skräddarsydda analysmoduler på de insamlade brottsplatsuppgifterna. Denna plattform innehåller ett lättanvänt gränssnitt som gör det smidigt att plugga in nya analysmoduler utan systemavbrott. På så vis kan polismyndigheten löpande utveckla nya analysmoduler och göra dessa tillgänglig för brottssamordnare m.fl. genom SAMS. I skrivande stund finns följande fyra analysmoduler implementerade i SAMS:
1. Prioritering av brott baserat på hur lika de är ett givet referensbrott.
2. Klustring av brott i olika grupper/kluster baserat på hur lika de är varandra.
3. Generering av beskrivande statistik över de insamlade brottsplatsuppgifterna för en valfri mängd brott.
4. Tidsanalyser för att undersöka hur en valfri mängd brott fördelar sig över tiden.

Ett exempel på hur en tidsanalys kan se ut återges i Figur 3 (a). Bilden visar resultatet för en aoristisk tidsanalys som kartlägger hur brottstiderna för 1 059 inbrott i centrala Malmö fördelar sig över såväl veckans olika dagar som timme på dygnet. Denna typ av analys kan vara användbara för t.ex. lokalpolisområdeschefer för att få en bild över hur brottsligheten inom olika brottskategorier fördelar sig i tiden. Det går snabbt att genomföra denna typ av analyser, för just detta exempel tog det 2 sekunder att slutföra analysen. Samma analys för 20 591 brott tar knappt 12 sekunder att slutföra. Att genomföra liknande analyser med ett mer manuellt arbetssätt tar dagar eller veckor.

Samtliga analyser i SAMS är mycket lättanvända eftersom de bygger på att man genomför analyserna på resultatet från en tidigare sökning bland brotten. Det betyder att brotten i ett sökresultat skickas vidare till valfri analysmetod. När analysmetoden är klar returneras analysresultaten till användaren. För några av analysmetoderna krävs även att man gör ett val för att precisera hur analysen ska genomföras. Ett exempel är för tidsanalyserna där man behöver ange vilken typ av tidsanalysmetod som ska användas; detta görs genom att välja en av de tillgängliga analystyperna i en rullgardinslista, se Figur 3 (b).

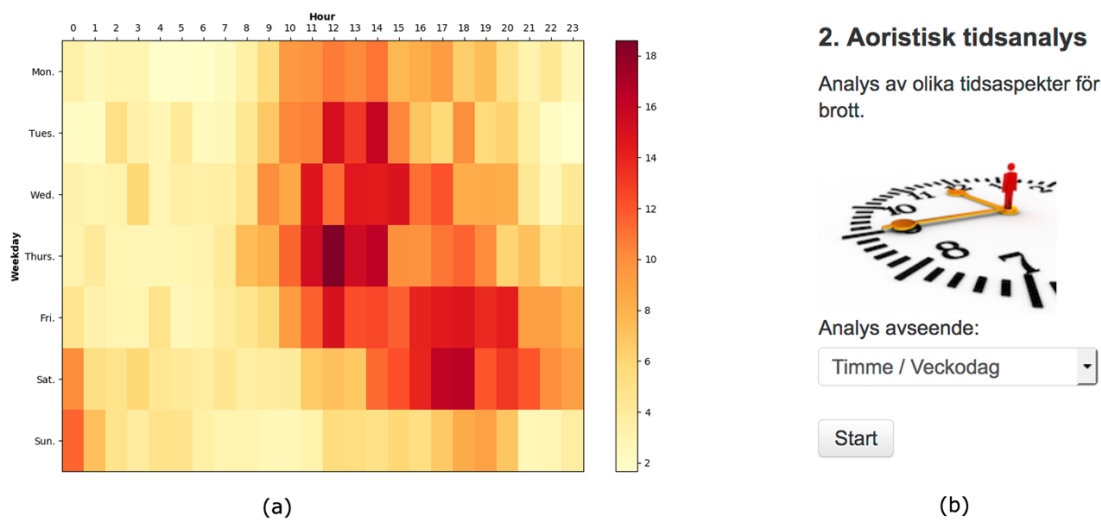

*Figur 3:Bilden (a) visar analysresultatet för en aoristisk tidsanalys i SAMS som kartlagt hur tiden för 1 059 inbrott fördelar sig över både veckodag och tid på dygnet. (b) visar hur den webb-baserade tidsanalysmetoden presenteras för användarna i SAMS.*

Det är även möjligt att automatisera båda sökningar och analysmetoderna vilket ger stora möjligheter. Genom att låta SAMS löpande i bakgrunden söka efter samband och mönster i strömmen av inkommande brottsplatsuppgifter ges möjligheten att identifiera likheter mellan nya brott och redan inrapporterade brott. Detta har alltså potential att kunna identifiera pågående serier av brott, t.ex. brottsturnéer utförda av så kallade MOCG:s (Mobile Organized Crime Groups) då en första digital brottsplatsbeskrivning kan komma in i systemet och bli sökbar redan minuter efter att en polispatrull besökt brottsplatsen. Att olika algoritmer automatiskt i bakgrunden kan analysera de inkomna strukturerade brottsplatsuppgifterna för att identifiera likheter i MO såväl som spatio-temporala samband ger värdefulla uppslag att jobba vidare med. På detta sätt kan relevanta brottsserier av brott utförda av gemensamma gärningsmän identifieras. Genom att utreda och lagföra dessa kan polisen därmed både öka personuppklarningsprocenten samtidigt de polisiära resurserna används mer effektivt.

## 6 Slutsatser

Denna initiala studie visar att den strukturerad metoden för registrering av bostadsinbrott är statistiskt signifikant mer tidseffektiv än traditionella fritextbaserade brottsanmälningar ($p < 0,05$), och att den senare är 66% långsammare. Samtidigt registrerar den strukturerade metoden signifikant fler brottsplatsuppgifter än fritextbaserade RAR-anmälningar ($p < 0,05$). Sammanvägt innebär detta att de strukturerade anmälningarna i snitt är 4,91 gånger mer effektiva än traditionella fritext-baserade anmälningar. Därtill finns betydande kvalitetsvinster med strukturerad brottsanmälningar jämfört med traditionella brottsanmälningar. Sammantaget finns det stora tidsbesparingar såväl som kvalitetsökningar att hämta hem för polismyndigheten inom exempelvis de s.k. vardagsbrotten för både yttre personal såväl som exempelvis polisens kontaktcenter (PKC). En grov uppskattning av besparingen landar någonstans i intervallet 11-30 heltidstjänster/år. Vilket skulle kunna avlasta redan tungt belastade poliser samt i viss mån frigöra resurser inom myndigheten som kan användas till mer kvalificerade uppgifter, t.ex. inom brottsamordningen. Samtidigt, erhålls mer enhetligt formaterade anmälningar av högre kvalitet och med signifikant fler relevanta brottsplatsuppgifter. Avslutningsvis möjliggör den strukturerade metoden även mer avancerade och träffsäkra analysmetoder baserat på de insamlade brottsplatsuppgifterna genom en fruktbar samverkan mellan polis och akademi.

## 7 Framtida arbete

Det finns flera uppslag till fortsatta studier baserat på resultaten i denna artikel. Till att börja med vore det intressant att genomföra en utökad studie som jämför mängden brottsplatsuppgifter i ett slumpvis urval av redan ifyllda skarpa RAR-anmälningar med SAB-formulär. Därutöver skulle det även vara intressant att utvärdera den strukturerade metoden för ytterligare brottskategorier, t.ex. för transport- och Dieselstölder som också har färdiga formulär utvecklade. Avslutningsvis skulle det också vara intressant att undersöka möjligheterna och eventuella risker med att samla in brottsplatsuppgifter med en interaktiv app istället för dagens PDF-formulär.

## 8 Acknowledgements